\documentclass{jnmp}

\usepackage{amsmath}

\usepackage{graphicx}
\usepackage{latexsym}

\JNMPnumberwithin{equation}{section}

\theoremstyle{definition}

\begin{document}

%
\renewcommand{\evenhead}{F Toppan}
\renewcommand{\oddhead}{On Anomalies in Classical Dynamical Systems}

%
\thispagestyle{empty}

\FirstPageHead{8}{4}{2001}{\pageref{toppan-firstpage}--\pageref{toppan-lastpage}}{Article}

\copyrightnote{2001}{F Toppan}

\Name{On Anomalies in Classical Dynamical Systems}
\label{toppan-firstpage}

\Author{Francesco TOPPAN}

\Address{CBPF - CCP, Rua Dr. Xavier Sigaud 150, cep 22290-180 Rio de Janeiro (RJ), Brazil\\
E-mail: toppan@cbpf.br}

\Date{Received February 27, 2001; Revised April 25, 2001; Accepted May
5, 2001}

\begin{abstract}
\noindent
The definition of ``classical anomaly" is introduced. It describes
the situation in which a purely classical dynamical system which
presents both a lagrangian and a hamiltonian formulation admits
symmetries of the action for which the Noether conserved charges,
endorsed with the Poisson bracket structure, close an algebra
which is just the centrally extended version of the original
symmetry algebra. The consistency conditions for this to occur are
derived. Explicit examples are given based on simple
two-dimensional models. Applications of the above scheme and lines
of further investigations are suggested.
\end{abstract}

\section{Introduction}

Here I define as ``{\em classically anomalous}" any classical
dynamical system, described both in the lagrangian and in the
hamiltonian formalism, whose symmetries of the action produce
conserved Noether charges which, under the Poisson-bracket
algebra, satisfy a centrally extended version of the original
symmetry algebra.

In many and perhaps even most of the texts
discussing quantum mechanics and quantum field theories it is
commonly stated that anomalies are a purely quantum-mechanical
effect. This statement reflects a widespread (but erroneous)
belief in the scientific community, mostly shared by researchers
who do not have a direct working experience with anomalies. While
the specialists in the field are aware that some specific
features, which can be reasonably named ``anomalies", can be
present even in purely classical dynamical systems, it seems,
however, that this correct interpretation passes largely
unnoticed. One of the reasons is due to the fact that most of the
results concerning anomalous effects in classical dynamical
systems are scattered in the literature. Moreover, they appear in
rather technical contexts and it seems that very little effort (if
any) has been made in order to place them in a more general
framework.

The aim of this paper is to furnish some
clarification, emphasizing one single aspect of the appearance of
``classical anomalies". In the interpretation proposed here
``classical anomalies", as previously defined, lay and can be
detected in the interplay between the
lagrangian-versus-hamiltonian description of a dynamical system
presenting a symmetry of the action. Very basic examples are
explicitly constructed and analyzed. No new result will be
discussed here, rather a re-interpretation of known results and
techniques will be given. Due to the mainly pedagogical character
of the present note, a minimum level of mathematical
sophistication has been purposely kept throughout the following
discussion.

 The present work is organized as follows. In the
next section a drastically sketchy and far from complete resume of
the history and importance of anomalies in quantum field theory
will be made. The result obtained by Gervais and Neveu~\cite{toppan:GN}
in analyzing the Liouville theory will be mentioned. To my
knowledge, they were the first authors who noticed an anomalous
effect in a classical system. These authors indeed observed,
according to the present definition, a ``classical anomaly".

In the following section a general argument is given suggesting
the mechanism which gives rise to what has been here named of {\em
classical anomalies}. The remaining sections are devoted to
working out some specific examples with concrete models, by
analyzing in each specific case whether the symmetries of the
actions are preserved or not by the Poisson bracket structure. For
simplicity reasons all the examples are worked out for
two-dimensional field theories. In order of presentation, the
following systems will be analyzed:
\begin{enumerate}
\vspace{-1.5mm}
\itemsep0mm
\item[{\em i)}] the free chiral fermion,
\item[{\em ii)}] the massless free boson,
\item[{\em iii)}] the Floreanini--Jackiw \cite{toppan:FJ} chiral boson (FJ) model (introduced to
complete the discussion of free and chiral models),
\item[{\em iv)}] finally, the Liouville theory will be revisited in light of the
present interpretation.
\vspace{-1.5mm}
\end{enumerate}

 In the conclusions some further
remarks and comments will be made and possible future lines of
development will be suggested.

\section{A bit of history}

In the last thirty years the investigation of anomalies in quantum
field theories has been one of the major areas of research
attracting the attention of theoretical physicists. The reason is
clear: an objective relevance for physical applications, coupled
with a beautiful mathematical structure.

Indeed, the first discovered anomaly (by Adler and Bell--Jackiw
\cite{toppan:Adl,toppan:BJ}, named ABJ after the authors) was the
$U(1)$ chiral anomaly of gauge theories putting a consistency
constraint on the existence of a quantized gauge theory. As an
important application, the gauge group and the representation
multiplets of the standard model for the electroweak interactions
are carefully selected in order to guarantee an anomaly-free
theory. Much in the same way, the critical dimensionality of a
quantized string theory can be determined by requiring the
cancellation of an anomaly~\cite{toppan:Sch} which can be
associated to the Weyl invariance~\cite{toppan:Pol}.

On the other hand, anomalies in physics are not always unwanted
features to be eradi\-ca\-ted. E.g. the trace anomalies associated to
dilatation invariance lead to the Callan--Symanzik equations~\cite{toppan:Cal}.
For a reference work concerning anomalies in the
context of quantum current algebras and their related physical
applications one can consult~\cite{toppan:Jac}.

Since their original
discovery, anomalies have been regarded as a feature of
quantization. Some folklore was put on this aspect. In the light
of the Feynman path-integral approach, Fujikawa~\cite{toppan:Fuj}
developed a celebrated method which relies upon the fact that the
functional measure is not always invariant under a symmetry of the
classical action.

On the mathematical side, anomalies have
been shown to satisfy consistency conditions induced by the
(anomalous) Ward identities satisfied by the corresponding quantum
field theories~\cite{toppan:WZ}. The (covariant) anomalies computed with
the Fujikawa method do not,however, satisfy such conditions;
nevertheless the relation between the two anomalies (consistent
versus covariant) was made explicit in~\cite{toppan:BZ}. The reason for
this discrepancy is that in the Fujikawa approach one regularizes
the jacobian arising from a field transformation instead of the
full partition function. A slightly modified version of the
Fujikawa technique can be introduced, allowing the regularization
of the full partition function. It turns out to be equivalent to
the heat kernel technique.

The \cite{toppan:WZ} consistency
conditions for the anomalies allow one to determine their
expressions via purely algebraic methods which make use of the
so-called ``transgression formula"~\cite{toppan:BCR}. An elegant
reformulation of such consistency conditions is given in terms of
the BRST-cohomology~\cite{toppan:BCR2}. To summarize, the {\em form} of
the possible anomalies is determined by the possible existence of
non-trivial cocycles for a BRST-cohomology associated to the
symmetry under consideration. On the other hand, the {\em
coefficients} of such anomalies can be related to the index
theorems for elliptic operators via heat kernel computations~\cite{toppan:Gil}
(for this purpose a one-loop Euclidean version of the
quantum field theory under investigation, regularized through
zeta-functions~\cite{toppan:HSC}, is required). A detailed account of the
latter construction is given in~\cite{toppan:Top}.

Anomalous symmetries of quantized theories have received,
therefore, a nice mathematical interpretation. As recalled, they
have been regarded as genuine new features of quantization, not
present in the underlining classical theories. This point of view
is, by the way, still commonly shared by most of the researchers
in the area and popularized in standard textbooks.

As far as I know, Gervais and Neveu were the first authors
(in~\cite{toppan:GN}, pages $131$/$2$) who observed in a classical
theory (in a purely quantum context such features had already been
observed in~\cite{toppan:CCJ}) a new phenomenon that, with some
good reasons, deserves to be named ``classical anomaly" and for
which the definition here introduced is applicable. The authors
of~\cite{toppan:GN} analyzed the Liouville theory appearing in the
partition function of non-critical strings according
to~\cite{toppan:Pol}. They showed that, even for the {\em
classical} Liouville theory, the generators of (one chiral sector
of) the conformal transformations satisfy, under classical Poisson
brackets, the Virasoro algebra, i.e., the central extension of the
Witt algebra. The remark appearing in~\cite{toppan:GN} was not
later developed (e.g, the Noether charges were not explicitly
mentioned), since the main focus of their authors was on the
quantum version of the Liouville theory (see also their related
works~\cite{toppan:GN2}). Needless to say, most of the papers
written by theoretical physicists on the Liouville field theory
deal with the quantum version of the model, as a simple key-word
inspection of the electronic bulletin boards reveals.

Subsequent works such
as~\cite{toppan:Bab} and~\cite{toppan:BBT} on classical Liouville and Toda-field
systems, were mostly concerned with the integrability properties
of such models, like the presence of classical Sklyanin
$r$-matrices in their Drinfeld--Sokolov exchange-algebras. No
connection of such classical Poisson-brackets structures with the
symmetries of the action, even if implicit, was explicitly
stated.

\newpage

\section{General considerations on classical anomalies}

The class of systems under consideration here consists of the
classical dynamical systems which admit both a lagrangian and a
hamiltonian description. It will be further assumed that the
action ${\cal S}$ admits an invariance under a group of symmetries
which can be continuous (Lie), infinite-dimensional and/or super.
The conserved Noether charges are associated to each generator of
the symmetries of the action. When the hamiltonian dynamics is
considered, the phase space of the theory possesses an algebraic
structure given by the Poisson brackets. The existence of such a
structure makes it possible to compute the Poisson bracket between
any two given Noether charges. In the standard situation, the
Poisson brackets among Noether charges realize a closed algebraic
structure which is isomorphic to the original algebra of the
symmetries of the action. It turns out, however, as it will be
illustrated in the examples which follow, that this is not always
the case. Indeed it can happen that the algebra of Noether charges
under Poisson bracket structure close a centrally extended version
of the original symmetry algebra. Mimicking the quantum case, the
following definition can be proposed for a classical dynamical
system. The system is said to possess an anomalously realized
symmetry, or in short a ``classical anomaly", if the following
condition is satisfied: the symmetry transformations of the action
admit Noether generators whose Poisson brackets algebra is a
centrally extended version of the algebra of symmetry
transformations.

Therefore a classical anomaly is a very
specific case of ``non-equivariant map" (for a discussion in a
finite-dimensional setting see~\cite{toppan:Mar}). Not all
non-equivariant maps discussed in the literature are classical
anomalies. For instance the one-dimensional free-particle
conserved quantities $p$ (the momentum) and $pt-mx$ generate a
non-equivariant map (the Poisson bracket between $p$ and $pt-mx$
is proportional to the mass $m$). However, despite being
conserved, they do not generate a symmetry of the action and for
that reason they are not Noether charges.

On the other hand,
infinite-dimensional non-equivariant moment maps were con\-struc\-ted
in~\cite{toppan:HK}. In those papers the only explicit application
concerned the dynamical systems of KdV type (classical integrable
hierarchies). Such systems, in contrast with the examples
discussed here, admits a hamiltonian description, but not a
lagrangian formulation. Even if conserved quantities can be
constructed, they can not be interpreted as Noether charges.

The possibility for a classical anomaly to occur is based on very
simple and nice mathe\-matical consistency conditions, implemented
by the Jacobi-identity property of the given symmetry algebra. Let
us illustrate this point by considering some generic (but not the
most general) scheme. Let us suppose that the (bosonic) generators
$\delta_a$'s of a symmetry invariance of the action satisfy a
linear algebra whose structure constants satisfy the Jacobi
identity, i.e.
\begin{equation*}
[\delta_a,\delta_b] ={f_{ab}}^c\delta_c,
\end{equation*}
while
\begin{equation*}
[\delta_a,[\delta_b,\delta_c]]
+[\delta_b,[\delta_c,\delta_a]] +[\delta_c,[\delta_a,\delta_b]]=0.
\end{equation*}
The associated Noether charges $Q_a$'s are further assumed to be
the generators of the algebra, i.e., applied on a given field
$\phi$ they produce
\begin{equation}
\delta_a\phi = \{Q_a,\phi\}, \label{noter1}
\end{equation}
where the brackets obviously denote the Poisson-brackets.

The condition
\begin{equation}
[\delta_a,\delta_b]\phi = {f_{ab}}^c\delta_c \phi,
\label{comm1}
\end{equation}
puts restriction on the possible Poisson brackets algebra
satisfied by the Noether charges. It is certainly true that
\begin{equation*}
\{ Q_a, Q_b\} = {f_{ab}}^c Q_c,
\end{equation*}
(which corresponds to the standard case) is consistent with both
(\ref{noter1}) and (\ref{comm1}). However, in a generic case, it
is not at all a necessary condition since more general solutions
can be found. Indeed, the presence of a central extension,
expressed through the relation
\begin{equation*}
\{ Q_a, Q_b\} = {f_{ab}}^c Q_c + k*\Delta_{ab},
\end{equation*}
(where $k$ is a central charge and the function $\Delta_{ab}$ is
antisymmetric in the exchange of $a$ and $b$), is allowed.

Indeed, since the relation
\begin{equation}
\{Q_a, \{Q_b,\phi\}\} - \{Q_b, \{Q_a,\phi\}\} =
\{\{Q_a,Q_b\},\phi\} \label{pb1}
\end{equation}
holds due to the Jacobi property of the Poisson bracket structure
(which is assumed to be satisfied), no contradiction can be found
with (\ref{comm1}); the right hand side of (\ref{pb1}) in fact is
given by
\begin{equation*}
 \{ {f_{ab}}^c Q_c + k*\Delta_{ab}, \phi\} = \{
{f_{ab}}^c Q_c, \phi\} = {f_{ab}}^c\delta_c\phi,
\end{equation*}
due to the fact that $k$ is a central term and has vanishing
Poisson brackets with any field.

 This observation on one hand puts
restrictions on the possible symmetries for which a classical
anomaly can be detected; the symmetries in question, on a purely
algebraic ground, must admit a central extension. This is not the
case, e.g., for the Lie groups of symmetry based on finite simple
Lie algebras. On the other hand one is warned that, whenever a
symmetry {\it does} admit an algebraically consistent central
extension, it should be carefully checked, for any specific
dynamical model which concretely realizes it, whether it is
satisfied exactly or anomalously. This remark already holds at the
classical level, not just for purely quantum theories.

Some further points deserve to be mentioned. The first one
concerns the fact that the quantization procedure (which, for the
cases we are here considering, can be understood as an explicit
realization of an abstract Poisson brackets algebra as an algebra
of commutators between operators acting on a given Hilbert space)
can induce anomalous terms for theories which, in their classical
version, are not anomalous in the sense previously specified. It
therefore turns out that the occurrence of classical anomalies is
a phenomenon which is ``more difficult to observe" than the
corresponding appearance of quantum anomalies since it occurs more
seldom.

A second point concerns the fact that the algebra of
Poisson brackets, as an abstract algebra, is assumed to satisfy
the Leibniz property. This is no longer the case for its concrete
realization given by the algebra of commutators. The Noether
charges are in general non-linearly constructed with the original
fields $\phi_i$ (which collectively denote the basic fields and
their conjugate momenta) of a given theory. For such a reason it
is only true in the classical case that, whenever an anomalous
central charge in an infinitesimal linear algebra of symmetries is
detected, it can be normalized at will by a simultaneous rescaling
of all the fields $\phi_i$ involved
($\phi_i\mapsto\alpha\cdot\phi_i$) and of the Poisson brackets
normalization ($\{.,.\}\mapsto \frac{1}{\alpha}
\{.,.\}$), for an arbitrary real constant $\alpha$. In the
classical case any central charge different from zero can
therefore be consistently set equal to $1$. However in the quantum
case a specific value of the central charge is fixed by the type
of representation of the symmetry algebra associated with the
given model and is a genuine physical parameter (the role of the
Virasoro central charge in labeling the conformal minimal models
is an example). The above argument is not, however, (at least
directly) applicable to non-linear symmetries, such as those
leading to the classical counterparts of the Fateev--Zamolodchikov
$W$-algebras. Classical non-linear symmetries fall outside the
scope of the present paper and deserve to be analyzed separately.

It should be noticed that the presence of a classical anomaly in
the construction of~\cite{toppan:IKP} is the underlining reason
which allows overcoming a no-go theorem and realizing a partial
breaking of an $N=4$ extended supersymmetry.

It is worth mentioning that in a different context, the appearance
of centrally extended algebras has been studied
in~\cite{toppan:KS} (and references therein). This analysis
however, developed for lagrangian dynamical systems, is not
directly related with the present results.

Furthermore, let me remark that the presence of a centrally
extended algebra of classical symmetries is not always a sign of
the presence of an anomaly (at least not in the sense specified
here). In~\cite{toppan:Kar} it was shown that a classical two-dimensional
complex bosonic field, coupled to an external constant
electromagnetic field, admits a symmetry corresponding to the
central extension of the two-dimensional Poincar\'e algebra. This
model is not anomalous, within the definition here proposed,
because, due to the presence of the constant external field, the
symmetry algebra of the classical action itself is centrally
extended and not given by the ordinary $2D$ Poincar\'e
algebra.

The term ``classical anomaly" has been employed
in~\cite{toppan:GP} as well, in a different context however and to
denote a different phenomenon than the one here discussed.

Finally, in the present work no effort is made to derive the
hamiltonian dynamics associated to a given lagrangian. It is
simply assumed to exist, based on the results furnished in the
literature. This is especially true for the chiral boson model in
Section~6, whose hamiltonian analysis is somewhat delicate, but
has nevertheless been performed in~\cite{toppan:Gir}, see
also~\cite{toppan:CDT}.

\section{The free chiral fermion}

The first example that will be discussed here concerns the theory
of the free chiral fermion. It is described by the Grassmann field
$\psi(x,t)$, where $x$ is a one-dimensional space coordinate and
$t$ the time. The dynamics is specified by the action
\begin{equation*}
{\cal S} = \int dxdt\cdot \psi\partial_-\psi,
\end{equation*}
where
\[
z_\pm =x\pm t
\qquad \mbox{and} \qquad
\partial_\pm =\frac{1}{2}(\partial_x\pm
\partial_t).
\]

For our purposes we will assume the space-coordinate $x$ to be
compactified on a circle $S^1$ of radius $R$ and $\psi$ to satisfy
periodic boundary conditions.

The equation of motion is given by
\begin{equation*}
\partial_-\psi = 0.
\end{equation*}
The action ${\cal S}$, besides being off-shell invariant under the
transformation specified by the infinitesimal function $\epsilon (
z_+)$
\begin{equation*}
\delta (\psi) = \epsilon(z_+)\partial_x\psi+\frac{1}{4}\epsilon'(z_+)\psi,
\end{equation*}
(the prime in the r.h.s. denotes the derivative)
admits a global
fermionic symmetry given by
\begin{equation*}
\delta_\kappa(\psi) = \kappa,
\end{equation*}
where $\kappa$ is a global fermionic parameter.

It is convenient to expand $\epsilon (z_+)$ as a Laurent series
according to
\begin{equation*}
\epsilon(z_+) = -\sum_n\epsilon_n(z_+)^{n+1}.
 \end{equation*}
The two above symmetries can be expressed through
\begin{gather*}
\delta (\psi) = \sum_n\epsilon_n \cdot l_n \psi, \qquad
\delta_\kappa (\psi)=  \kappa\cdot g\psi,
\end{gather*}
where the operators $l_n$ and $g$ are respectively given by
\begin{gather*}
l_n = -(z_+)^{n+1} \partial_x-\frac{1}{2}(n+1)(z_+)^n,\qquad
g= \oint dx \cdot \frac{\delta}{\delta\psi(x)} .
\end{gather*}
While the commutators among the $l_n$'s operators realize the Witt
algebra
\begin{equation*}
[ l_n, l_m] = (n-m) l_{n+m},
\end{equation*}
the anticommutator of $g$
with itself satisfies
\begin{equation*}
[g,g]_+ = 0,
\end{equation*}
so that $g$ is nilpotent.

The conserved Noether charges
associated to the above symmetries are given by
\begin{gather*}
L_n = -\oint dx (z_+)^{n+1} \psi\partial_x\psi,\qquad
 G= 2 \oint dx\psi(x,t).
\end{gather*}
In the hamiltonian description the equation of motion is expressed
through
\begin{equation*}
{\dot \psi} = \{ H, \psi\}_t, 
\end{equation*}
where $H$ is the hamiltonian
\begin{equation*}
H = -\oint dx\cdot (\psi\partial_x\psi),
\end{equation*}
while the equal-time Poisson brackets $\{.,.\}_t$ are introduced
through
\begin{equation*}
\{\psi (x),\psi(y)\}_t = \frac{1}{2}\delta(x-y).
\end{equation*}
We are now in the position to compute the Poisson brackets among
the Noether conserved charges, which are the generators of the
symmetries, according to
\begin{gather*}
l_n\psi = \{ L_n, \psi\}_t,\qquad
g\psi = \{ G, \psi\}_t.
 \end{gather*}
While the Noether generators $L_n$ associated with the Witt
generators reproduce, under the Poisson brackets structure, the
Witt algebra, i.e.
\begin{equation*}
\{ L_n, L_m\} = (n-m) L_{n+m},
\end{equation*}
it is no longer true that the fermionic conserved
charge $G$ satisfies the same nilpotency condition as $g$. Indeed
we have that the Poisson bracket of $G$ with itself produces a
central element, given by
\begin{equation*}
\{G,G\} = 4\pi R.
\end{equation*}
It follows that the generator of the fermionic symmetry is now no
longer nilpotent. Even in this trivial free model the presence of
a symmetry which presents a classical anomaly can be detected.

In the quantum case, due to the double contractions in the Wick
expansion, the quantum analogs of the $L_n$ generators satisfy the
centrally extended version of the Witt algebra, i.e. the Virasoro
algebra, with central charge $c=\frac{1}{2}$. This is in
accordance with the statement that the ``quantization" is a more
effective way to produce anomalies than the plain introduction of
a classical Poisson bracket structure. Still, as the fermionic
symmetry shows, in many cases the introduction of a classical
Poisson bracket structure is sufficient to induce anomalies at the
level of the Noether charges.

\section{The free massless boson in \mbox{\mathversion{bold}$2D$}}

The next example that we would like to discuss concerns the
$2$-dimensional free massless boson model, described by the
following action
\begin{equation*}
{\cal S} = -2\int dx dt \cdot\partial_-\phi\partial_+\phi.
\end{equation*}
The field $\phi(x,t)$ satisfies the free equation of motion
\begin{equation*}
\Box \phi\equiv 4\partial_-\partial_+\phi = 0.
\end{equation*}
This system admits an (anomalous-free) two-dimensional conformal
invariance which corresponds to the direct sum of two copies of
the Witt algebra ($Witt\oplus Witt$). Actually the symmetry
algebra of the system is richer. Indeed, the following
transformations are symmetries of the action
\begin{gather*}
\delta_+\phi = \epsilon(z_+)\partial_+\phi +\mu(z_+),\qquad
\delta_-\phi = {\overline\epsilon}(z_-)\partial_-\phi
+{\overline\mu}(z_-),
\end{gather*}
for arbitrary infinitesimal functions
$\epsilon(z_+)$, ${\overline\epsilon}(z_-)$, $\mu(z_+)$,
${\overline{\mu}}(z_-)$. Such a set of transformations is anomalous
in the sense discussed here. This point can be easily understood
when we consider a specific case of $\mu(z_+)$
(${\overline{\mu}}(z_-)$) given by
\begin{gather*}
\mu(z_+) = \lambda_+\partial_+\epsilon(z_+),\qquad
{\overline{\mu}}(z_-)=\lambda_-\partial_-{\overline{\epsilon}}(z_-).
\end{gather*}
for arbitrary fixed values of the parameters $\lambda_\pm$.

In full analogy with the previous case, after Laurent series
expansion for $\epsilon (z_+)$, ${\overline{\epsilon}}(z_-)$,
\begin{gather*}
\epsilon (z_+) = -\sum_n\epsilon_n (z_+)^{n+1},\qquad
{\overline{\epsilon}}(z_-) = -\sum_n{\overline{\epsilon}}_n(z_-)^{n+1},
\end{gather*}
we obtain two mutually commuting set of $\lambda_+$ and
$\lambda_-$-dependent symmetry generators, each set generating a
copy of the Witt algebra. They are given by
\begin{gather*}
l_n(\lambda_+)=
-{(z_+)}^{n+1}\partial_+-\lambda_+(n+1){(z_+)}^n\cdot\int
\frac{\delta}{\delta\phi(x,t)},\nonumber\\
\overline{l}_n(\lambda_-)=
-{(z_-)}^{n+1}\partial_--\lambda_-(n+1){(z_-)}^n\cdot\int
\frac{\delta}{\delta\phi(x,t)}.
\end{gather*}
For any given couple of values $\lambda_\pm$, we obtain the
closure of the $Witt\oplus Witt$ algebra
\begin{gather*}
[l_n(\lambda_+),l_m(\lambda_+)] =
(n-m)l_{n+m}(\lambda_+),\nonumber\\
[\overline{l}_n(\lambda_-),\overline{l}_m(\lambda_-)] =
(n-m){\overline l}_{n+m}(\lambda_-),\nonumber\\
[l_n(\lambda_+),\overline{l}_m(\lambda_-)] = 0.
\end{gather*}
The free massless boson model admits a hamiltonian formulation,
with the hamiltonian given by
\begin{equation*}
H = \frac{1}{2}\int dx \left(\pi^2+(\partial_x\phi)^2\right).
\end{equation*}
The equations of motions, expressed through
\begin{equation*}
\frac{d}{dt}f=\{H,f\} + \frac{\partial}{\partial t} f
\end{equation*}
imply
\begin{gather*}
{\dot\phi} =\pi,\qquad
 {\dot\pi}=(\partial_x)^2\phi.
\end{gather*}
The equal-time Poisson brackets are obviously given by
\begin{equation*}
\{\pi(x),\phi(y)\}= \delta (x-y)
\end{equation*}
and vanishing otherwise.

As a straightforward computation
shows, the conserved Noether charges $L_n (\lambda_+)$,
\linebreak ${\overline
L}_n(\lambda_-)$, associated to the symmetry generators
$l_n(\lambda_+)$, ${\overline l}_n(\lambda_-)$ respectively, are
recovered from the Laurent expansions
\begin{gather*}
L_n(\lambda_+) = \int dx (z_+)^{n+1}\cdot T,\qquad
{\overline L}_n(\lambda_-) =\int dx (z_-)^{n+1}\cdot {\overline T},
\end{gather*}
where $T$, ${\overline T}$ are given by
\begin{gather*}
T =\frac{1}{4}(\pi^2+(\partial_x\phi)^2+2\pi\partial_x\phi -
4\lambda_+{\partial_x}^2\phi-4\lambda_+\partial_x\pi),\nonumber\\
{\overline T} =-\frac{1}{4}(\pi^2+(\partial_x\phi)^2-2\pi\partial_x\phi
-4\lambda_-{\partial_x}^2\phi+4\lambda_-\partial_x\pi).
\end{gather*}
The conservation law for $L_n$, ${\overline L}_n$ is a consequence
of the (anti-)chiral equations satisfied by $T$ (${\overline T}$)
respectively, i.e.
\begin{gather*}
 \partial_- T = 0,\qquad
\partial_+ {\overline T}= 0.
\end{gather*}
$L_n$, ${\overline
L}_n$ are the generators of the $l_n$, ${\overline l}_n$
transformations since
\begin{gather}
l_n\phi = \{L_n,\phi\},\qquad
{\overline l}_n\phi = \{{\overline L}_n, \phi\}. \label{llbar}
\end{gather}
 $L_n$, ${\overline L}_n$ generate the direct sum of two
copies of the Virasoro algebra, $Vir\oplus Vir$, as can be
directly read from the equal-time Poisson brackets between
$T(x)$, ${\overline T}(x)$, namely
\begin{gather*}
\{ T(x), T(y)\} = -2{\lambda_+}^2{\partial_y}^3\delta(x-y) +2
T(y)\partial_y\delta(x-y) + \partial_y
T(y)\cdot\delta(x-y),\nonumber\\
\{T(x),{\overline T}(y)\}= 0,\nonumber\\
\{ {\overline T}(x), {\overline T}(y)\} =
2{\lambda_-}^2{\partial_y}^3\delta(x-y) +2 {\overline
T}(y)\partial_y\delta(x-y) + \partial_y {\overline
T}(y)\cdot\delta(x-y).
\end{gather*}
For given values of $\lambda_\pm\neq 0$, central terms are
produced which are proportional to ${\lambda_\pm}^2$. The
corresponding transformations can therefore be regarded as
anomalous.

The two-dimensional conformal symmetry itself however is not
anomalous in this free case, since for the choice $\lambda_\pm
=0$, the symmetry is preserved at the Poisson bracket level.

\renewcommand{\footnoterule}{\vspace*{3pt}%
\noindent
\rule{.4\columnwidth}{0.4pt}\vspace*{6pt}}

It should be stressed the fact that the freedom in choosing
inhomogeneous transformations acting on $\phi$, for
$\lambda_\pm\neq 0$, can be held as  responsible for the
preservation (i.e. not anomalous realization) of the
two-dimensional conformal invariance even in the quantum case. The
choice $\lambda_\pm\neq 0$ corresponds to the introduction of the
Feigin--Fuchs term in the Coulomb gas formalism\footnote{Let me
recall that in the quantum OPE language, given a chiral propagator
$\phi(z)\phi(w)\sim - \log(z-w)$, a stress-energy tensor $T(z)$
satisfying a Virasoro algebra can be introduced through
$T(z)=-\frac{1}{4}:\partial\phi\partial\phi: +
i\alpha\partial^2\phi$. The linear term in $\phi (z)$ is inserted
in order to allow modifying the value of the central charge $c$ of
the Virasoro algebra, given by $c= 1 - 24\alpha^2$. This
construction can be repeated in the classical case too.}.

\section{The Floreanini--Jackiw chiral boson model}

For completeness, let us discuss the last chiral and free model,
namely the Floreanini--Jackiw chiral boson model~\cite{toppan:FJ}
introduced through the lagrangian
\begin{equation*}
{\cal L} = \partial_t\phi\partial_x\phi -(\partial_x\phi)^2,
\end{equation*}
which leads to the equation of motion
\begin{equation*}
\partial_x\partial_-\phi=0.
\end{equation*}
Despite the fact that it is not manifestly Lorentz-invariant, it
can nevertheless be shown to be Poincar\'{e} invariant in $2$
dimensions. This model defines the dynamics of a chiral boson. The
treatment is much in the same lines as the free boson model with a
notable exception. Since we are in presence of chiral dynamics the
invariance of the model is given by a single (chiral) copy of the
Witt algebra and its central extension. A class of
$\lambda$-dependent infinitesimal symmetries of the above action
is given by
\begin{equation}
\delta_\lambda \phi =
\epsilon(z_+)\partial_x\phi +\lambda\partial_x\epsilon (z_+) .
\label{transfo}
\end{equation}
The corresponding
 Noether conserved charges are given by the following
expressions
\begin{equation}
L_n (\lambda)= \frac{1}{2}\int dx
(z_+)^{n+1}\left((\partial_x\phi)^2+\lambda{\partial_x}^2\phi\right).
\label{not}
\end{equation}
The hamiltonian of the system is
\begin{equation*}
H= \frac{1}{2}\int dx (\partial_x\phi)^2,
\end{equation*}
while the
Poisson-brackets structure  system is non-local
\begin{equation*}
\{\phi(x),\phi(y)\}={\partial_y}^{-1}\delta(x-y).
\end{equation*}
Despite its non-locality however, since in (\ref{not}) only the
derivatives of the field $\phi$ enter, the algebra satisfied by
the $L_n(\lambda)$ Noether charges is a local algebra which, as in
the previous example, corresponds to the Virasoro algebra with
central extension proportional to $\lambda^2$. For $\lambda\neq 0$
we are in the presence of an anomaly induced by the Poisson
bracket structure. The Noether charges $L_n$ are, as in the
previous example, the generators of the transformations in~(\ref{transfo}).

\section{The Liouville theory revisited}

The last model that we are going to discuss is the Liouville
theory, revisited in view of the considerations which motivated
the present paper.

The action of the Liouville model can be
written as
\begin{equation*}
{\cal S}= -\int dx dt\cdot(2 \partial_-\phi\partial_+\phi
+e^{2\phi}).
\end{equation*}
The equation of motion is
\begin{equation*}
2\partial_-\partial_+\phi = e^{2\phi}.
\end{equation*}
In the hamiltonian description the hamiltonian is given by
\begin{equation*}
H = \oint dx \cdot
\left(\frac{1}{2}\pi^2+\frac{1}{2}(\partial_x\phi)^2 + e^{2\phi}\right),
\end{equation*}
while the Poisson-bracket structure between $\pi$, $\phi$ is the
same as in the free case
\begin{equation*}
\{\pi(x),\phi(y)\}=\delta(x-y).
\end{equation*}
We obtain
\begin{gather*}
{\dot \phi}= \pi,\qquad
{\dot \pi}= {\partial_x}^2\phi -2 e^{2\phi}.
\end{gather*}
The theory is conformally invariant, with transformations given,
as in the free-case, by the infinitesimal transformations
\begin{gather*}
\delta_+\phi = \epsilon(z_+) \partial_x\phi +
\lambda_+(\partial_x\epsilon(z_+))\phi,\qquad
\delta_-\phi = {\overline\epsilon}(z_-)\partial_x\phi
+\lambda_-(\partial_x{\overline\epsilon}(z_-))\phi .
\end{gather*}
However, due to the presence of the potential term, the action is
no longer off-shell invariant for arbitrary values of
$\lambda_\pm$. The invariance is indeed satisfied only for
\begin{equation*}
\lambda_+=\lambda_-=\frac{1}{2}.
\end{equation*}
There is no longer a whole class of $\lambda_\pm$-dependent
symmetry transformations, but just a given, point-like in the
$\lambda_\pm$ parametric space, set of symmetry transformations.
In this particular case, the analysis plainly follows the one
conducted for the free massless boson. The conserved Noether
charges $L_n$, ${\overline L}_n$ can be introduced through the
Laurent-expansion of
\begin{equation*}
\int dx \epsilon(z_+)
T(x),\qquad  \int dx {\overline\epsilon} (z_-) {\overline T}(x).
\end{equation*}
They are conserved provided that
\begin{equation*}
\partial_- T=\partial_+{\overline{T}}=0.
\end{equation*}
$T$, ${\overline{T}}$ can be unambiguously fixed to be given by
\begin{gather*}
T=\frac{1}{4}\left(\pi^2+(\partial_x\phi)^2+2\pi\partial_x\phi
+e^{2\phi} -2{\partial_x}^2\phi-2\partial_x\pi\right),\nonumber\\
{\overline T} =-\frac{1}{4}\left(\pi^2+(\partial_x\phi)^2-2\pi\partial_x\phi
+e^{2\phi} -2{\partial_x}^2\phi+2\partial_x\pi\right).
\end{gather*}
The two sets of Noether charges, $L_n$, ${\overline L}_n$, generate
the symmetry transformation of the field $\phi$ according to~(\ref{llbar}).

Their Poisson-bracket algebra however is
anomalous and coincides with $Vir\oplus Vir$, with fixed values of
the two central charges $c_\pm$ given by $c_\pm = \mp 6$, for the
given normalization of the field $\phi$ and of the action,
\begin{gather*}
\{ T(x), T(y)\} = -\frac{1}{2}{\partial_y}^3\delta(x-y) +2
T(y)\partial_y\delta(x-y) + \partial_y
T(y)\cdot\delta(x-y),\nonumber\\ \{T(x),{\overline
T}(y)\}=0,\nonumber\\ \{ {\overline T}(x), {\overline T}(y)\} =
\frac{1}{2}{\partial_y}^3\delta(x-y) +2 {\overline
T}(y)\partial_y\delta(x-y) + \partial_y {\overline
T}(y)\cdot\delta(x-y).
\end{gather*}
(I recall that, by definition, the central charge is normalized to
be the coefficient in front of the inhomogeneous term
${\delta}'''$ normalized by a factor $12$). The conformal
invariance of the $2D$ Liouville theory is classically anomalous,
satisfying the definition proposed here. This is in contrast with
the free massless boson model, where the symmetry can be restored
both at the classical and quantum level, as well as the free
chiral fermion theory. In that case the chiral (i.e. Witt)
invariance is classically preserved, while it is violated at the
quantum level for a fixed value of the central charge ($c=1/2$).
The Liouville theory, on the other hand, admits a non-vanishing
classical central charge. Its normalization is meaningless in the
classical case, since it can always be reabsorbed through a
simultaneous rescaling of the fields and of the Poisson brackets,
as previously mentioned. Nevertheless, in the quantum theory, the
effect of such ``freedom of rescaling" of the underlining
classical theory can be seen in the arbitrariness of the Liouville
quantum central charge, which is not fixed by the theory, apart
the restriction coming from unitarity requirement. This is in
sharp contrast to the free chiral theory, where such freedom is
not allowed.

\section{Conclusions}

In the present work I have stressed the fact that features which
correspond to an anomalous realization of a symmetry can be
present even in purely classical dynamical systems.

I introduced the definition of ``classical anomaly" to describe
the situation of a classical system whose conserved Noether
charges, which are associated to a symmetry of the action, admit a
Poisson brackets algebra, induced by the hamiltonian dynamics,
which is only isomorphic to a centrally extended version of the
original symmetry algebra. A~classical anomaly is therefore a
specific example of a non-equivariant moment map realized by
Noether charges.

The underlining mathematical reason which makes possible the
realization of such a case has been illustrated in Section~3. It
has been shown there that eventual central extensions of the
original symmetry realized by the Poisson brackets algebra of the
Noether charges satisfy the compatibility constraint required by
the simultaneous validity of~(\ref{noter1}) and~(\ref{comm1}).
Later, a list of simple models which show the concrete application
of this mechanism and the appearance of anomalies have been
given.

 It is certainly true, as discussed in the text, that
quantization is still the ``preferred" mechanism to produce
anomalies (a good example is given by the chiral fermion theory of
Section~4, which is quantum anomalous under $1$-dimensional
diffeomorphisms, but sa\-tis\-fies the ordinary Witt algebra for what
concerns classical Poisson brackets). It turns out, however, that
in general and in many cases of interest, it is not necessary to
perform the quantization of a dynamical system in order to induce
anomalies. In some cases the introduction of a classical Poisson
brackets structure is sufficient for the purpose. The anomalous
nilpotent fermionic symmetry of the free chiral fermion of Section~4
 is perhaps the simplest example, as well as the anomalous
conformal symmetry of the classical Liouville equation analyzed in
Section~7.

Moreover, any symmetry which algebraically admits
a central extension, is potentially anomalous. The investigation
of the Poisson brackets algebra of its Noether charges realized on
specific models can lead to non-trivial results even for classical
dynamics.

Specific differences with respect to the ``quantum
anomalies" have been pointed out throughout the text. At least for
the case of symmetries associated to linear algebras (in the
present analysis no effort was put in including infinitesimal
symmetries of non-linear $W$-type), the Leibniz rule observed by
classical Poisson brackets allow, through the simultaneous
rescaling of the fields and the Poisson brackets, to freely
normalize the value of the central charge, which can be
conveniently chosen.

The examples chosen and the techniques
employed in the present work are elementary. The main motivation
of this paper is to illustrate, in the simplest possible contexts,
the mathematical framework, deep and simple at the same time,
behind the appearance of anomalies in classical dynamical
systems.

The techniques which are usually encountered in the
literature and which appear in disguised form in the analysis of
the examples here illustrated (e.g., the introduction of the
Feigin--Fuchs term in the Coulomb gas approach to ``shift" the
value of the quantum central charge), often appear to a layman
reader as just a set of ad hoc prescriptions to perform technical
computations. While it is certainly true that they are technically
very helpful, the deep symmetry principles which make them
possible are somehow hidden. To place them in the proper context
of lagrangian and hamiltonian dynamics is the main issue of the
present paper.

The analysis here conducted suggests many
possible lines of development. On a purely mathematical ground one
can ask which kind of centrally extended algebras can find a
dynamical interpretation as (anomalous) symmetry for some given
dynamical system.

D. Leites \cite{toppan:DL} has recently proposed a
very specific problem on the extended superization of the
Liouville equation which could be investigated in the light of the
present considerations.

On the other hand, the interplay
between lagrangian and hamiltonian methods seems quite fruitful.
It seems likely that, by employing superspace techniques, the
embedding of certain classes of hamiltonian solitonic equations in
some superized system, which also admits a (super)lagrangian
description, could be given. This subject is currently under
investigation. Nice and neat results concerning the symmetry
algebra of these systems should be derivable. Needless to say, the
presence of central charges in the Virasoro subalgebra is
mandatory for any integrable system which contains KdV as its
consistent reduction.

\label{toppan-lastpage}
\end{document}